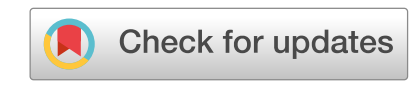

OPEN

# High-quality superconducting α-Ta film sputtered on the heated silicon substrate

Yanfu Wu[1], Zengqian Ding[1], Kanglin Xiong[1,2]✉ & Jiagui Feng[1,2]✉

**Intrigued by the discovery of the long lifetime in the α-Ta/$Al_2O_3$-based Transmon qubit, researchers recently found α-Ta film is a promising platform for fabricating multi-qubits with long coherence time. To meet the requirements for integrating superconducting quantum circuits, the ideal method is to grow α-Ta film on a silicon substrate compatible with industrial manufacturing. Here we report the α-Ta film sputter-grown on Si (100) with a low-loss superconducting $TiN_x$ buffer layer. The α-Ta film with a large growth temperature window has a good crystalline character. The superconducting critical transition temperature ($T_c$) and residual resistivity ratio (RRR) in the α-Ta film grown at 500 °C are higher than that in the α-Ta film grown at room temperature (RT). These results provide crucial experimental clues toward understanding the connection between the superconductivity and the materials' properties in the α-Ta film and open a new route for producing a high-quality α-Ta film on silicon substrate for future industrial superconducting quantum computers.**

Superconducting materials grown on Si or $Al_2O_3$ substrates may form films of high crystallinity and with inherently low dielectric loss, thus are explored as the materials for building superconducting quantum circuits[1–6]. Recently, researchers have made efforts to search for new superconducting films with stable superconducting properties and mature processing used in the quantum computing field, aiming to improve the performance of superconducting qubits, including long coherence time and fast gates[5,7–12]. The high-quality superconducting films with low dielectric losses at surfaces and interfaces which exhibit high RRR are promising for the fabrication of high-performance qubits[1,2,7,13,14]. Specially, using α-Ta films to fabricate the 2D Transmons, the devices have shown a significant improvement in performance arising from the lower surface-related loss[7,8]. Thus, the α-Ta film is a promising base superconductor to build large-scale superconducting quantum circuits with high-performance properties, paving the way toward practical superconducting quantum computers. However, in these superconducting qubits studies[7,8], the sapphire substrate which was used to grow α-Ta film cannot be easily scaled-up with advanced integration, such as the through via technology. By contrast, the silicon substrate is widely used for large-scale integrated circuits. It is therefore very natural to raise the question of whether α-Ta film can be grown on a silicon substrate or not.

The obtainment of α-Ta film which is easily formed at high temperature deposited on Si substrate without inner-diffusion interface is very limited, partially because of the obstacle that Ta is highly reactive to the heating Si substrate[15–17]. Although it has been reported that the α-Ta film is deposited on Si substrate at RT successfully by using several strategies such as optimizing the sputtering conditions and adding under layers[18–30]. In comparison to high-temperature growth, these films are more likely to have smaller grain sizes, more grain boundaries, and more surface defects as a result of RT deposition[18–22], which might lead to an additional dielectric loss in the superconducting quantum device[8,12–14,31–33]. Besides, in these studies, the Ta-Si interface may include thicker non-superconducting underlayers[25,27] or metal silicides[15–17] which might form because of heating treatments used during the device fabrication flow. This would increase microwave loss channels at interfaces[12,13,31–33]. Thus, we need a new method to grow α-Ta film on the Si substrate that has a large grain size and clear interface with low-loss superconducting buffer layers while minimizing dielectric loss at surfaces and interfaces to improve superconducting qubit performance.

Here, we systematically investigated the quality and superconductivity of α-Ta film prepared on a superconducting $TiN_x$ buffer layer deposited on Si (100) substrate. Previous studies have demonstrated that a significant improvement in the performance of superconducting quantum circuits can be achieved by using $TiN_x$ as the base superconductor in the capacitor and microwave resonators, illustrating $TiN_x$ film have a low dielectric loss[34,35]. α-Ta films were formed at different temperatures varied from RT to 500 °C, directly suggesting a large processing temperature window. This is in sharp contrast to previous studies[19–23], in which the preparation of α-Ta film was under RT conditions. The crystal quality of the α-Ta film was improved by increasing the growth temperature. Meanwhile, the difference in the growing temperature has no effect either on the low intensity of contaminants in the α-Ta films or on the clean and sharp interface between Si and Ta due to the presence of the $TiN_x$ buffer layer. Furthermore, we observed the zero resistance and measured RRR in the α-Ta film. Notably, RRR found in α-Ta films grown at 500 °C is remarkably higher than that in α-Ta grown at RT, which possibly be attributed to the large grain size and the suppression of surface defects.

[1]Gusu Laboratory of Materials, Suzhou 215123, China. [2]Suzhou Institute of Nano-Tech and Nano-Bionics, CAS, Suzhou 215123, China. ✉email: klxiong2008@sinano.ac.cn; jgfeng2017@sinano.ac.cn

## Methods

α-Ta possesses a cubic structure with the lattice parameter a = 0.33 nm[19,20]. To synthesize high-quality α-Ta films on high resistivity 2 inches Si(100) substrates (its resistivity value > 10 kΩ cm), a two-step method was applied. First, the low-loss superconducting $TiN_x$ buffer layer, 3~5 nm thick was deposited by dc reactive magnetron sputtering using 2 inches Ti (purity of 99.995%) target and $N_2$ (purity of 99.999%) reactive gas. After cleaning the substrates with wet chemicals (see "**Wet chemical processes of Si substrates**" in the Supplementary Information), they were thermally cleaned inside the growth chamber at 500 °C for 30 min. Then, the substrates were cooled down to room temperature at 30 °C per minute. During the $TiN_x$ deposition, the substrate temperature was held at RT, while a constant pressure of 2 mTorr was maintained in the presence of Ar and $N_2$, flowing at 10 sccm and 15 sccm respectively. The power of the DC generator was 100 W. After $TiN_x$ of deposition, the α-Ta films were prepared on the $TiN_x$ buffer layers under different temperatures while the Ar pressure was kept at 5.25 mTorr with a gas flow of 20 sccm, and the power of the DC generator was 200 W.

## Results and discussions

The crystal structure and phases of the α-Ta film were analyzed using X-Ray Diffraction (XRD). Figure 1 shows the XRD of α-Ta films grown on $TiN_x$ buffer layers (The information about the $TiN_x$ buffer layer is in the Supplementary Information) at different temperatures. It can be cleanly seen that the dominative features of α-Ta films are the (110) and (220) diffraction peaks near 38.1° and 81.5° respectively. In addition, the weak α-Ta (100) peaks at 54.8° are visible [Fig. 1b]. Apart from strong dominant diffraction peaks of α-Ta, only a very tiny diffraction peak of β-Ta is observed at 34.6°, suggesting within a wide temperature range, a major α-phase Ta film on Si (100) substrate was deposited successfully. This is most likely because of $TiN_x$ buffer layer promoted the growth of α-Ta film due to its reducing lattice mismatch[19,20]. Furthermore, as the temperature increased, not only the relative intensity of the main (110) peak became stronger, also its full width at half maximum (FWHM) became sharper. This measurement result is direct evidence for the grain size increase with increasing temperature, which is consistent with the previous report[20], indicating a high temperature led to the good crystallization of α-Ta film.

The quality of the sample surface was investigated with Atomic Force Microscope (AFM). The surface morphologies of α-Ta films grown on $TiN_x$ buffer layers at RT- 500 °C temperature range are revealed in typical large-scale AFM images (Fig. 2). At RT, the small round grains can be seen on the surface. With the substrate temperature increasing, elongated-like grains start to be obvious, as shown in Fig. 2b. However, the surface [Fig. 2c] of the α-Ta film deposited at 300 °C is visually different. A typical needle-shape feature of the grains is evident. At 400 °C, a tight network of elongated needle-shaped grains uniformly distributed across the surface [Fig. 2d]. Compared to the one sputter deposited at 300 °C, the size of the needle-shaped grains at 400 °C is wider and more homogeneous. Next at 500 °C, as shown in Fig. 2e, the grains look elongated, highly homogeneous in size and uniformly distributed on the surface. Moreover, the density of defects such as holes with varying size and depth in the surface of α-Ta grown at RT are much higher than these of α-Ta films grown at high temperatures. Figure 2f–g reveal that α-Ta films deposited at RT and 500 °C are highly dense and not porous, which suggests holes observed from AFM images only appear on the surface. These results reflect substrate temperature plays a critical role in controlling the surface morphology of the α-Ta film. The difference in the topographic feature is answerable for the different quality of macroscopic electrical transport properties in the films as discussed below.

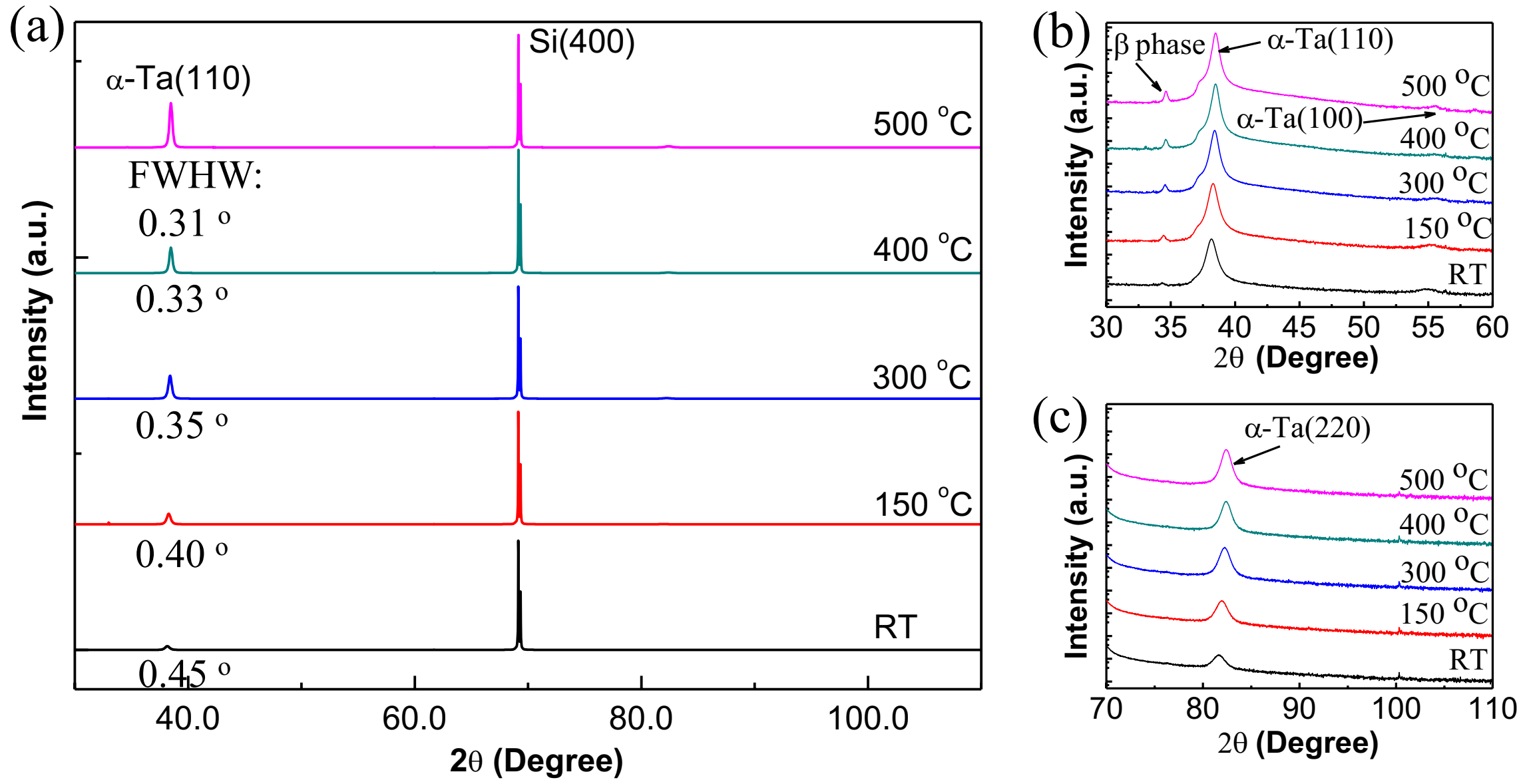

**Figure 1.** (**a**) XRD pattern of α-Ta films deposited on $TiN_x$/Si at different temperatures ranging from RT to 500 °C. The curves are shifted upward for a better display. (**b**) and (**c**) Enlarged XRD pattern near (110) peaks and (220) peaks shown in (**a**). The intensities are plotted on a logarithmic scale to see the important behavior at lower intensities shown in (**b**) and (**c**).

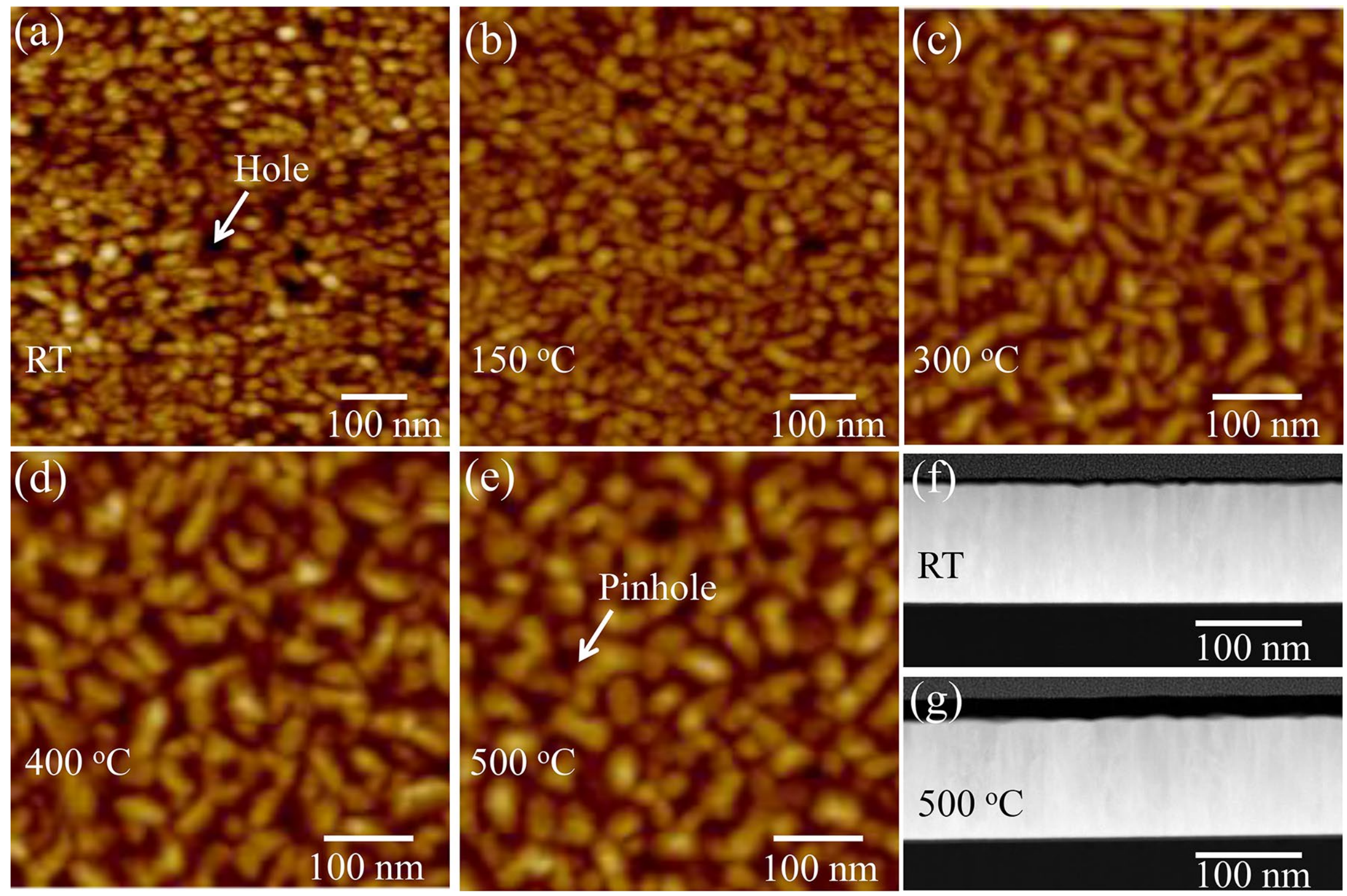

**Figure 2.** AFM images and cross section TEM of α-Ta/Si at different temperatures. (**a**, **f**) RT, (**b**) 150 °C, (**c**) 300 °C, (**d**) 400 °C, (**e**, **f**) 500 °C. White arrows are used for guiding holes in (**a**) and (**e**).

Close microstructure analysis near the hetero-interface between α-Ta film and Si by high-resolution high-angle annular dark-field scanning transmission electron microscopy (HAADF-STEM) and Energy Dispersive Spectrometer (EDS) is shown in Fig. 3. For α-Ta film grown at RT, the HAADF-STEM micrograph revealed that there is about 5 nm $TiN_x$ buffer layer between the Ta film and the Si substrate [Fig. 3a]. Additionally, the $TiN_x$ buffer layer growth on the bare Si substrates is correlated with the formation of an amorphous layer (1~2 nm thick) during the pre-deposition procedure, which is consistent with previous reports[36] (Supplementary Information Fig. S3). It has been reported that $TiN_x$ film can provide low dielectric loss in superconducting quantum computing systems[34,35]. And so it is reasonable to consider that if using α-Ta film deposited on the $TiN_x$ buffer layer as a material platform to build superconducting qubits, the $TiN_x$ buffer layer would not lead to additional dielectric loss. The EDS elemental maps of Si and Ta, as shown in Fig. 3b,c, exhibit step functions of the chemistry changes across the interfaces, illustrating that no intermixing of Ta and Si occurred between the α-Ta film and Si substrate. With respect to α-Ta film grown at 500 °C, similarly, Si and Ta' signals sharply change across the Ta/Si interface as shown in Fig. 3e and f. It should be concluded that no inter-diffusion in the Ta/Si interface at 500 °C, which is different from previous studies[15–17]. This result suggests the $TiN_x$ buffer layer is thermodynamically stable and dense, and therefore could prevent intermixing of Ta and Si between α-Ta film and Si substrate under high substrate temperature. It is noted that the backgrounds of EDS maps in these two samples are different, which might correlate to the different concentrations of contaminants introduced during the SEM sample preparation process.

The time of flight secondary ion mass spectroscopy (TOF-SIMS) results of the Ta films deposited at RT and 500 °C are shown in Fig. 4. Here, in order to directly compare the relative concentrations between the samples, all the signals are normalized by the intensity of Ta in the Ta layer region. These two samples both have almost the same Si, Ta, $TiN_x$, and $TaO_x$ profiles with sharp changes in the Ta/Si interfaces, showing no intermixing of Ta, $TiN_x$, Ti, and Si components. The result is directly evidenced that even heating to 500 °C, the Ta/$TiN_x$ and $TiN_x$/Si interfaces in the α-Ta film are as sharp and clean as that in the α-Ta film grown at RT, indicating the versatility of 5 nm thick $TiN_x$ buffer layer as discussed above. Meanwhile, the distribution of contaminants such as H, C and O in the bulk of films of these two samples are detected. As shown in Fig. 4c and d, similar O, C and H profiles are also observed. In addition, their concentrations are just above the detection levels. Based on this discussion, we can conclude that the insignificant intensity of contaminants is nearly the same in the bulk of these two films.

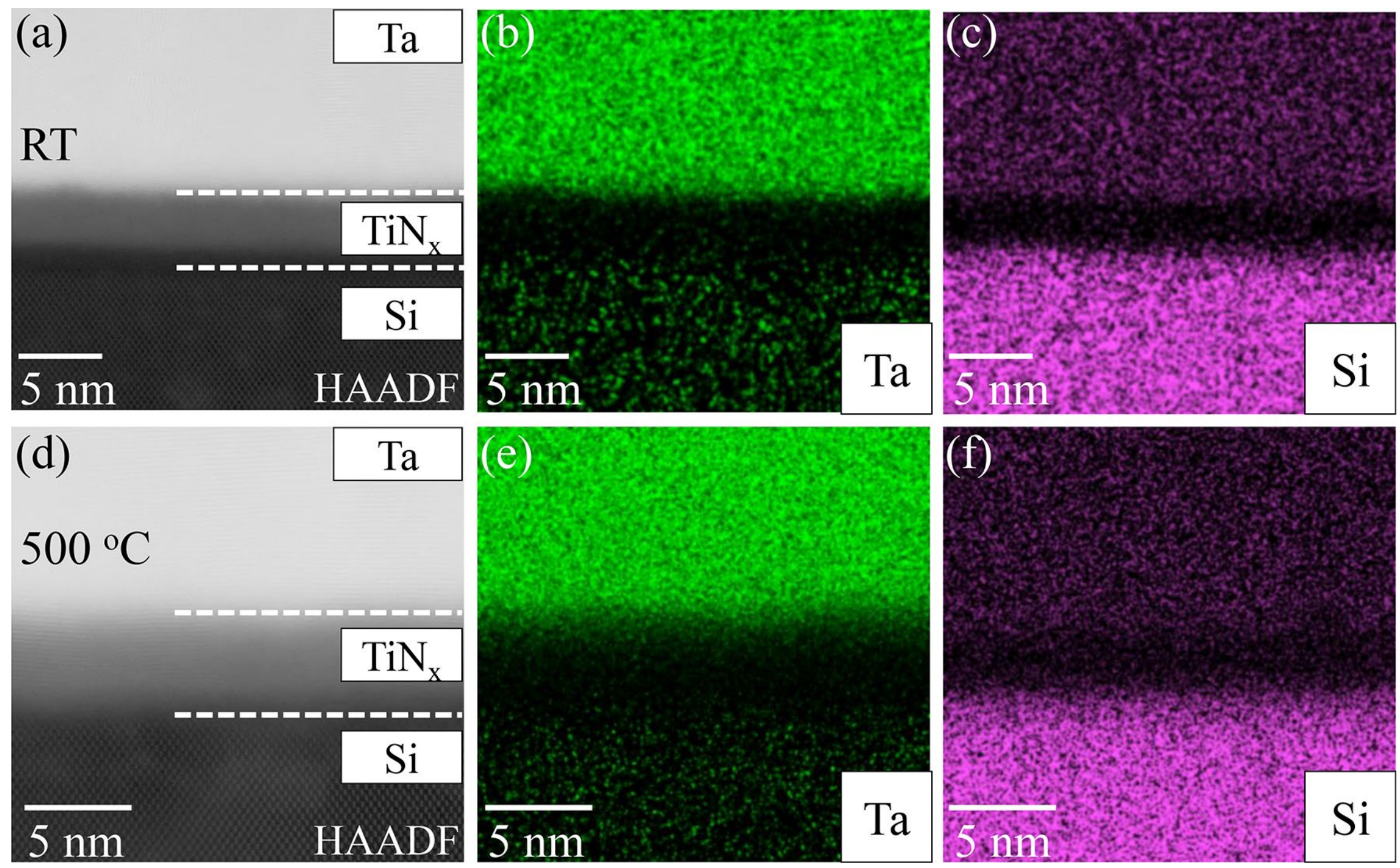

**Figure 3.** Ta-Si interface of α-Ta/Si films grown at RT (**a**–**c**) and 500 °C (**d**–**f**). (**a**, **d**) High resolution HAADF-STEM images showing Ta-Si interface; (**b**–**c**, **e**–**f**) EDS maps of Ta and Si.

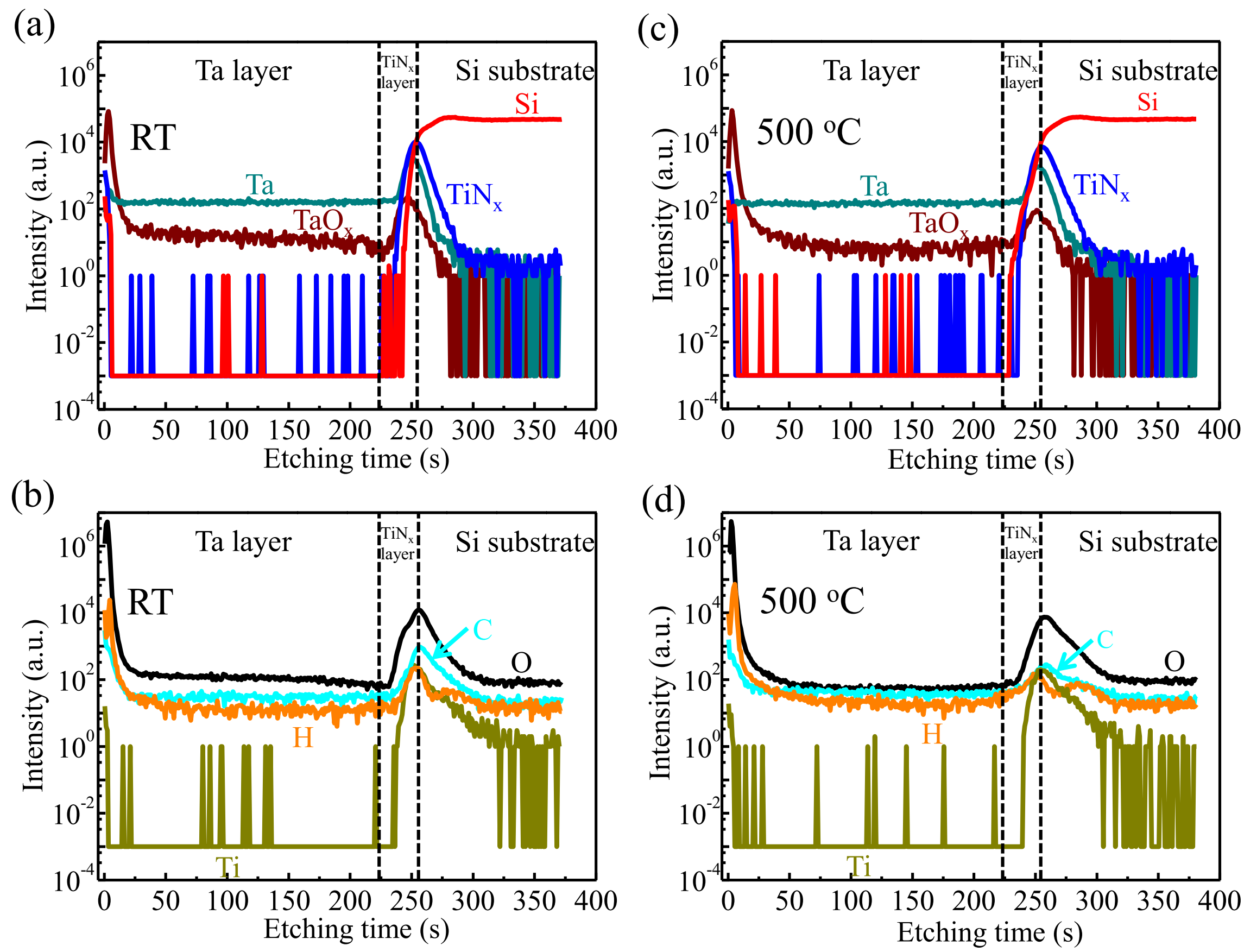

**Figure 4.** TOF-SIMS depth profiles of α-Ta/Si films deposited at (**a**, **b**) RT and (**c**, **d**) 500 °C respectively. The intensities are plotted on a logarithmic scale to see the important behavior at lower intensities.

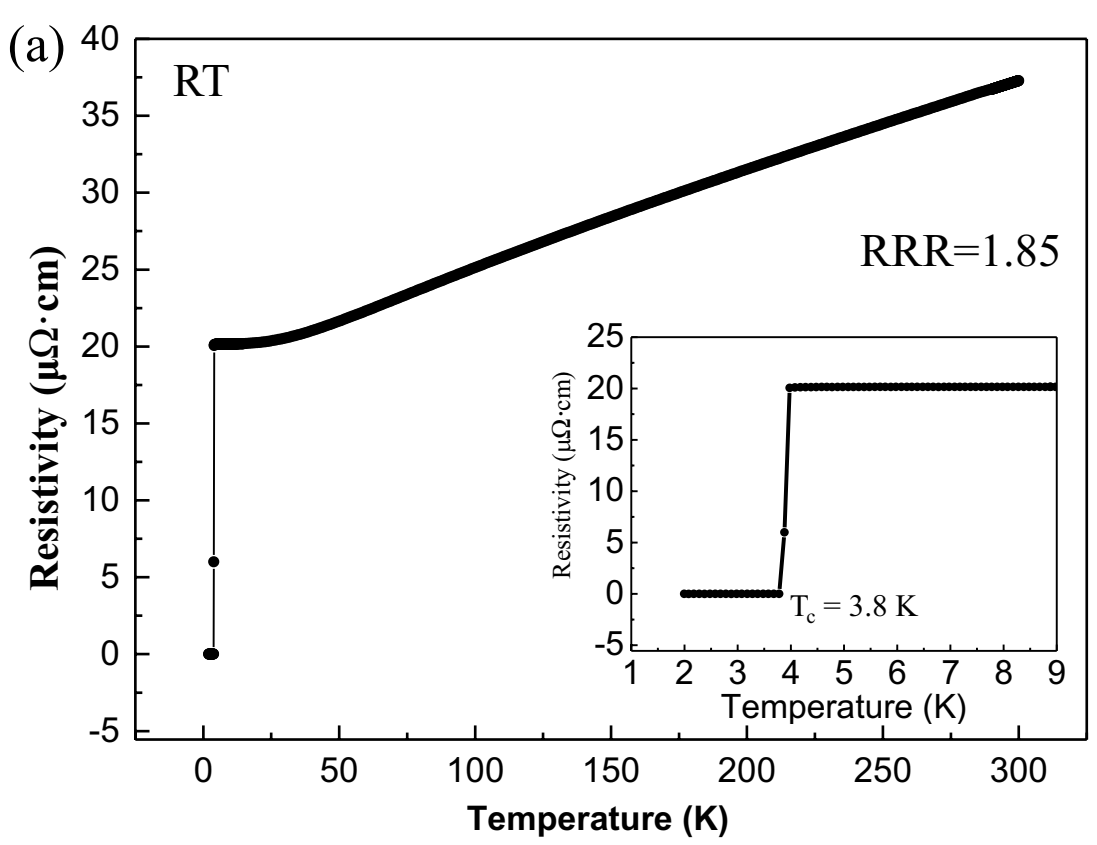

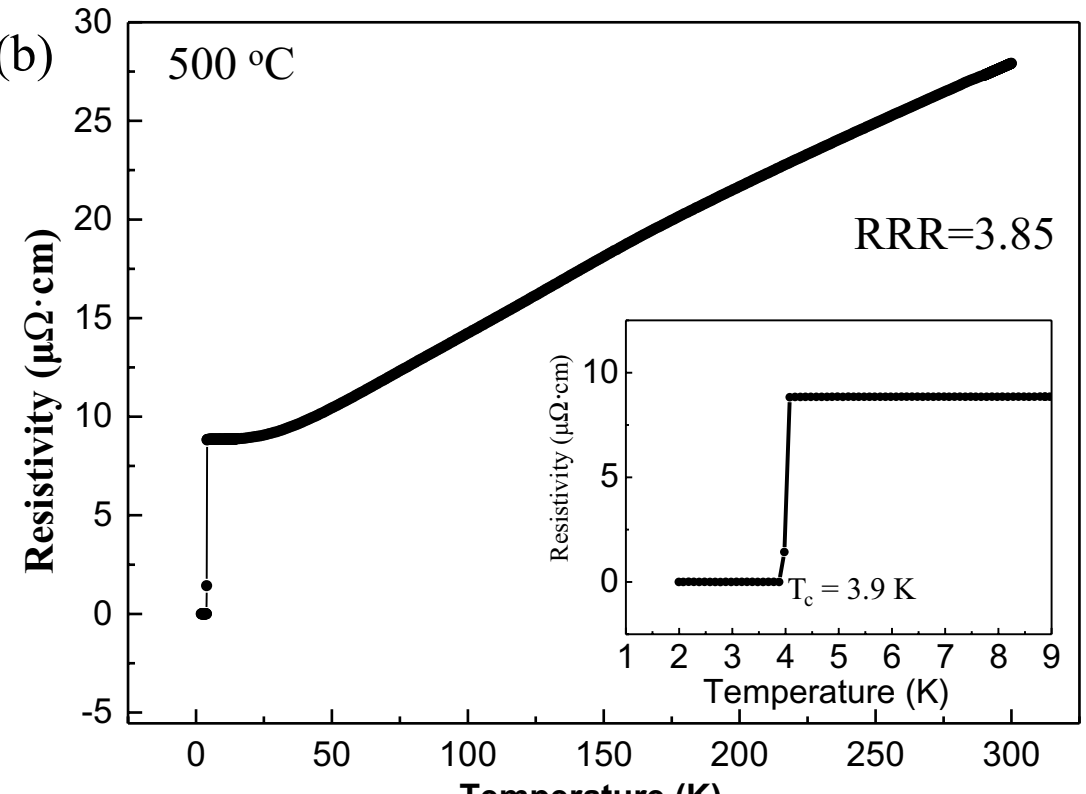

**Figure 5.** Temperature dependence of the electrical transport properties for α-Ta films deposited at (**a**) RT and (**b**) 500 °C respectively. The inset shows R-T curve near $T_c$.

The electrical properties were detected from the two α-Ta films grown at RT and 500 °C with four-point probe measurements. The resistivity measured as a function of temperature is shown in Fig. 5. The $T_c$ values are found to be 3.8 and 3.9 K for samples grown at RT and at 500 °C respectively. These Tc values are comparable to that of the bulk α-Ta but much higher than that of the bulk β-Ta ($T_c < 1$), directly verifying the deposited Ta film is in the α phase, which is consistent with the above XRD results. For the α-Ta film grown at 500 °C, its RRR value ($RRR = \rho(300\ K)/\rho(5\ K)$) is 3.85. Compared to α-Ta film grown at RT, where RRR = 1.85, the RRR value of α-Ta film grown at 500 °C is much higher and the resistivity much smaller. This is likely due to the larger grain sizes and the existence of fewer defects[37–40] rather than the presence of contaminants in the α-Ta film grown at 500 °C, which have been discussed above. Moreover, according to earlier studies[13,39,40], a higher RRR in the α-Ta film grown at 500 °C render it more suitable as a base material for superconducting multi-qubits.

## Conclusions

In summary, the α-Ta film deposited on Si(100) substrate with low-loss superconducting $TiN_x$ buffer layer has been studied comprehensively. The XRD results and AFM surface images show that good crystallization of α-Ta film with fewer surface defects was achieved by substrate temperature optimization. The identical low impurity concentrations and similar sharp Ta/Si interfaces with no inter-diffusion at various substrate temperatures indicate that the low-loss superconducting $TiN_x$ buffer layer is the dominant factor in the growth of the α-Ta film. The $T_c$ and RRR values in the α-Ta films are revealed by analyzing R-T curves. The increase in $T_c$ and RRR values is associated with the increase in grain size and the decrease in surface defects, providing a new hint to the correlation between the nature of superconductivity and the quality of the α-Ta film. Our result here shows that by manipulating the film growth, α-Ta film on Si(100) substrate with a sharp interface can be synthesized in both high quality and strong superconducting state, thus making it suitable to be used in large-scale superconducting qubit devices.

## Data availability

The data that support the findings of this study are available from the corresponding author upon reasonable request.

## Acknowledgements

K. L. X acknowledges support from the Youth Innovation Promotion Association of Chinese Academy of Sciences (2019319). J. G. F. acknowledges support from the Start-up foundation of Suzhou Institute of Nano-Tech and Nano-Bionics, CAS, Suzhou 20 (Y9AAD110).

## Author contributions

Y.W.: Investigation, Data analysis, Writing – original draft. Z.D.: Investigation, Data analysis. K.X.: Project administration, Writing – review & editing. J.F.: Project administration, Conceptualization, Supervision, Writing – review & editing.

## Competing interests
The authors declare no competing interests.

## Additional information
**Supplementary Information** The online version contains supplementary material available at https://doi.org/10.1038/s41598-023-39420-y.

**Correspondence** and requests for materials should be addressed to K.X. or J.F.